# Long Range Force Transmission in Fibrous Matrices Enabled by Tension-Driven Alignment of Fibers

Hailong Wang[1], Abhilash Nair[1], Christopher S. Chen[2], Rebecca G. Wells[3] and Vivek B. Shenoy[1*]

[1]Department of Materials Science and Engineering, [3]Departments of Medicine (GI) and Pathology and Laboratory Medicine, University of Pennsylvania, Philadelphia, PA 19104 and [2]Department of Biomedical Engineering, Boston University, Boston MA 02215

**Abstract:** Cells can sense and respond to mechanical signals over relatively long distances across fibrous extracellular matrices. Recently proposed models suggest that long-range force transmission can be attributed to the nonlinear elasticity or fibrous nature of collagen matrices, yet the mechanism whereby fibers align remains unknown. Moreover, cell shape and anisotropy of cellular contraction are not considered in existing models, although recent experiments have shown that they play crucial roles. Here, we explore all of the key factors that influence long-range force transmission in cell-populated collagen matrices: alignment of collagen fibers, responses to applied force, strain stiffening properties of the aligned fibers, aspect ratios of the cells, and the polarization of cellular contraction. A constitutive law accounting for mechanically-driven collagen fiber reorientation is proposed. We systematically investigate the range of collagen fiber alignment using both finite element simulations and analytical calculations. Our results show that tension-driven collagen fiber alignment plays a crucial role in force transmission. Small critical stretch for fiber alignment, large fiber stiffness and fiber strain-hardening behavior enable long-range interaction. Furthermore, the range of collagen fiber alignment for elliptical cells with polarized contraction is much larger than that for spherical cells with diagonal contraction. A phase diagram showing the range of force transmission as a function of cell shape and polarization and matrix properties is presented. Our results are in good agreement with recent experiments, and highlight the factors that influence long-range force transmission, in particular tension-driven alignment of fibers. Our work has important relevance to biological processes including development, cancer metastasis and wound healing, suggesting conditions whereby cells communicate over long distances.

## 1. Introduction

Cells in fibrous matrices sense and respond to mechanical forces over distances many times their diameter. Although cells cultured on polyacrylamide gels fail to sense substrate stiffness or the presence of other cells beyond a distance of about 20-25 μm (1–3), long-range force sensing (250–1000 μm) between cells in fibrous gels has been appreciated for decades. Stopak and Harris and later Miron-Mendoza et al. placed fibroblast explants into collagen gels and observed collagen realignment parallel to the connecting axes between explants, with translocation of collagen fibrils towards the explants, shortening of the axes, and fibroblast migration across the newly-aligned collagen fibril bridges (4, 5). Others have shown that single cells as well as cell colonies are able to align and compact collagen fibers over long distances (6, 7) and that these aligned fibers are required for long-range cell-cell interactions (7, 8). More recently, Winer et al. showed that single cells in fibrin gels were able to stiffen the gels both locally and globally (9).

---

[*] Corresponding Author: vshenoy@seas.upenn.edu

Long-range force transmission has significant relevance in normal physiology and pathophysiology over a range of length scales. At the level of single cells, mechanically-based cell-cell communication over long distances regulates patterning, including both tube formation and the detachment of cells from multicellular aggregates (7, 9, 10). At the tissue level, long-range force transmission may drive the development of tendons, ligaments, and muscle (4); it has the potential to mediate other large-scale architectural rearrangements typical of developmental processes as well (11). Long-distance force transmission between groups of cells, or cells and the matrix, may also mediate tissue-scale rearrangements in pathological settings such as pulmonary fibrosis and liver cirrhosis (12, 13). There are some experimental data implicating it in cancer metastasis (7, 14, 15), although other work suggests caveats to these findings (16).

Previous studies attempting to explain the mechanism of long-range force transmission have implicated applied strain and the presence of a fibrous network (6). While some investigators suggest that the strain-hardening properties of fibrous materials could explain long-range mechanical communication (9, 17) more recent evidence (8, 18) suggests that the fibrous nature of the ECM, specifically the presence of cross-linked fibers (primarily collagen) is critical in order for force to be transmitted over scales that are 10-20 times the diameters of the cells. Ma et al. used microscopy images to develop finite element models that included fibers that bridge pairs of interacting cells in a collagen matrix (8). They found that including discrete fibers along with a non-linear strain hardening matrix leads to long range transmission of forces, with the fibers carrying most of the loads and non-linear and isotropic matrix mechanics playing a relatively minor role. In other words, the fibrous nature of the collagen matrix, rather than a non-linear response to force, determined the extent of force transmission. It should be noted that since the fiber distribution in the model of Ma et al. was obtained from experiments, the model cannot predict how an initially random fiber network under strain yields reinforcing fibrous structures in response to forces from contractile cells. Multi-scale finite element models, where discrete fiber networks are employed to determine forces at nodes, have also been used to study force transmission in fibrous gels (18). It has also been observed that the shapes of cells play a crucial role in the transmission of forces. Fabry and coworkers reported that invasive tumor cells are elongated and spindle shaped compared to their non-invasive counterparts and they observed through displacements of beads in the matrices that force transmission is much longer ranged in the former than in the latter case (19). Elongated cells have also been found to be polarized (*i.e.* the forces they exert are aligned with their long axes). Although these and other studies (20, 21) have considered the role of individual cell and matrix elements in force transmission, none have addressed in an integrated way the impact of fiber realignment, the shape of the cells, the anisotropy and the magnitude of the contractile forces and the mechanical properties of fibrous gels on the long-range nature of force transmission.

In this work we develop a new non-linear and anisotropic constitutive description of fibrous materials that accounts for the long-range force transmission. We incorporate the fact that these fibrous materials stiffen preferentially along the directions of tensile principal stretches. We start from random and isotropic distributions of fibers, and from there study how mechanical anisotropy evolves as loads are applied. We have developed a finite element implementation of this constitutive law and have used it to study interactions of cells in 3D matrices and on fibrous substrates. In the case of simple cell geometries (spheres, ellipsoids, polarized vs. non polarized), we solve for the stress fields by analytic methods. Thus, we describe here an approach to systematically determine the role of fiber alignment, non-linear elasticity of fibers, cell shape, and polarization of contraction in long-range force transmission. We show that collagen fiber alignment is critical and that anisotropy in cell shape and contraction result in significantly greater collagen alignment and force transmission.

## 2. A New Constitutive Law for Fibrous Matrices

We first developed a new constitutive law to explain the behavior of fibrous matrices and to serve as the foundation for further simulations examining the impact of cells and their contractility on these matrices. To start, we carried out discrete fiber simulations (see Appendix A). We assume that when a fibrous matrix undergoes stretch, there are two families of fibers: the set of fibers that align with the direction of the maximum principal stretch as the material is loaded (fibers colored red in Fig. 1b) and the set of fibers that do not align with the applied load and thus display an isotropic mechanical response. When we plot stress vs. strain for such collagen networks (Fig. 1c), we find that there is a "knee" in the curve representing strain stiffening. This "knee," which according to our simulations requires the presence of the two families of fibers, is in good accord with experimental data (experiment, Ref (22) and Fig. 1d). For strains below a typical threshold (typically 5-10%, depending on collagen concentration and crosslinking), the network shows a nearly isotropic response, without stiffening. Beyond this threshold, there is a transition to a stiffening response that is concomitant with the formation of aligned fibers in the direction of maximum tensile stretch. With increased loading, the numbers of these highly aligned fibers increase, leading to the observed stiffening and to the alignment shown in insets in Fig. 1a and Fig. 1b.

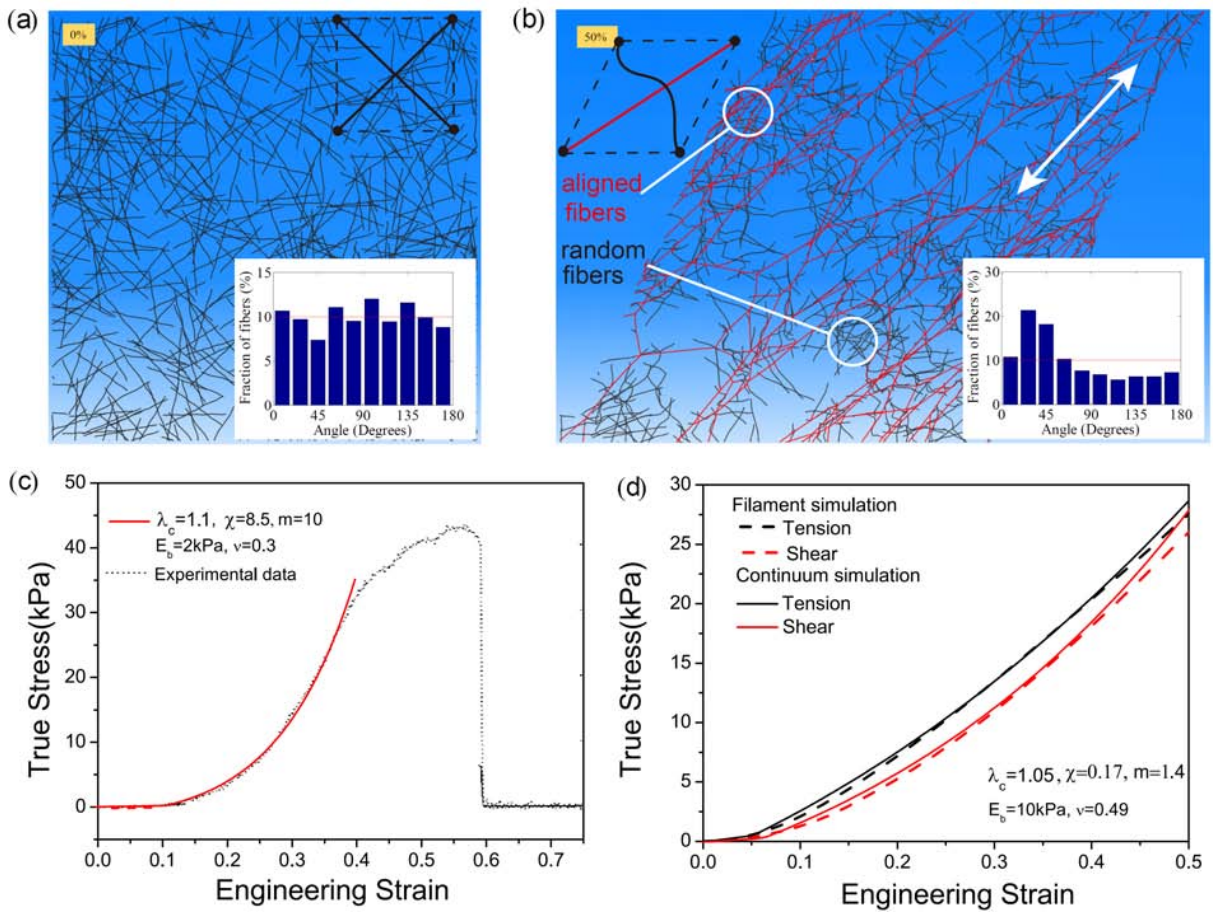

Figure 1: (a)-(b) Discrete fiber simulations of a random fiber network before (a) and after (b) shear deformation (50% shear strain). Insets show that the initial random distributions of fibers (a) develop a peak close to the 45° orientation (b), which coincides with the direction of maximum principle stretch. Fibers (with axial strain > 1%) that reorient along the tensile loading axis are colored in red. The white arrow in (b) indicates the direction of principle tensile stretch. (c) The stress-strain curves of collagen I under uniaxial deformation derived experimentally [Ref (22)] (black) are in good accord with those predicted from our constitutive law (red). The "knee" indicates strain stiffening at strains around 10%. The material parameters that provide the best fit to the experimental data are $\lambda_c = 1.1, \chi = (1+v)(1-2v)E_f/(1-v)E_b, E_b = 8.5, m = 10, E_b = 2\text{kPa}, v = 0.3$. (d) Stress-strain curves under uniaxial tension (black) and shear (red) deformations from discrete fiber simulations and from our constitutive law. The material parameters that provide the best fit to the discrete simulations are $\lambda_c = 1.05, \chi = 0.17, m = 1.4, E_b = 10\text{kPa}, v = 0.49$.

To capture the presence of these two distinct families of aligned and isotropic fibers when developing our constitutive law, we assume that the overall energy density $W$ of the collagen network consists of two contributions(23),

$$W = W_b + W_f$$

$$W_b = \frac{\mu}{2}(\bar{I}_1 - 3) + \frac{\kappa}{2}(J - 1)^2 \tag{1}$$

$$W_f = \sum_{a=1}^{3} f(\lambda_a)$$

Here the first term $W_b(\bar{I}_1, J)$ captures the isotropic response, which we describe using the neo-Hookean hyperelastic model, where $\kappa$ and $\mu$ are initial bulk and shear moduli, respectively and $W_f$ is the contribution from the aligned fibers. In the above equation, $F_{ij} = \partial x_i/\partial X_j$ is the deformation gradient tensor, where $X$ and $x$ labeled the reference and deformed coordinates respectively and $C = F^T F$ and $B = F F^T$ are the right and left Cauchy–Green deformation tensor, respectively. The invariants $J, C$ and $B$ can be defined as (23),

$$J = \det(F) \qquad C = \sum_{a=1}^{3} \lambda_a^2 N_a \otimes N_a \qquad B = \sum_{a=1}^{3} \lambda_a^2 n_a \otimes n_a \tag{2}$$

where $\lambda_1, \lambda_2, \lambda_3$ are the principle stretches, $\bar{I}_1$ is the first invariant of the deviatoric part of $C$ and $N_a$ and $n_a$ are the unit vectors in the principle stretch orientations in the reference state and deformed state, respectively. The functional form $f(\lambda_a)$ is chosen such that the system stiffens only in the direction of tensile principal stretches (beyond a critical value of tensile stretch as observed in experiments and discrete fiber simulations). This is accomplished by decomposing the Cauchy stress (true stress), $\sigma$, into isotropic ($\sigma^b$) and fibrous contributions ($\sigma^f$) (23),

$$\sigma = 2F \cdot (\partial W/\partial C) \cdot F^T/J,$$

$$\sigma = \sigma^b + \sigma^f$$

$$\sigma^b = \kappa(J - 1)I + \mu \text{dev}(\bar{B})/J \tag{3}$$

$$\sigma^f = \frac{1}{J}\sum_{a=1}^{3}\frac{\partial f(\lambda_a)}{\partial \lambda_a}\lambda_a(\mathbf{n}_a\otimes\mathbf{n}_a),$$

where $\mathbf{I}$ is the identity tensor and $\bar{\mathbf{B}} = \mathbf{B}/J^{2/3}$ is the left modified Cauchy–Green tensor. The principal components of the filamentous contribution can be obtained from

$$\frac{\partial f(\lambda_a)}{\partial \lambda_a} = \begin{cases} 0, & \lambda_a < \lambda_1 \\ \dfrac{E_f(\lambda_2 - \lambda_1)\left(\dfrac{\lambda_a - \lambda_1}{\lambda_2 - \lambda_1}\right)^{n+1}}{n+1}, & \lambda_1 \leq \lambda_a < \lambda_2 \\ E_f\left[\dfrac{\lambda_2 - \lambda_1}{n+1} + \dfrac{(1+\lambda_a-\lambda_2)^{m+1}-1}{m+1}\right], & \lambda_a \geq \lambda_2, \end{cases} \quad (4)$$

chosen such that the principal stresses vanish below the critical (tensile) principal stretch, $\lambda_c$ and show a stiffened response characterized by the modulus $E_f$ and a strain hardening exponent, $m$. In order to ensure that the derivative of the stress-strain curve is continuous near the transition point, a smooth interpolation function is used between $(\lambda_c - \lambda_t/2, \lambda_c + \lambda_t/2)$, where the transition width, $\lambda_t = 0.25\lambda_c$, transition exponent $n = 5$, and we have defined $\lambda_1 = \lambda_c - \lambda_t/2$, $\lambda_2 = \lambda_c + \lambda_t/2$. The functional form $f(\lambda_a)$ which leads to Eq. 4 is provided in Appendix B.

*Biophysical basis for the constitutive law*: The strain energy function and the stresses we propose depend on two parameters (the initial bulk and shear moduli) for the isotropic response and three parameters (the critical stretch, $\lambda_c$, the initial modulus of the fibrous phase, $E_f$ and the strain hardening exponent of the fibrous phase, $m$) for the anisotropic response. We have determined these parameters for collagen networks by comparing the stress-strain curves for uniaxial and shear deformation from discrete network simulations with our constitutive model (Fig. 1d). The biophysical basis that underlies the constitutive law that we have postulated is the presence of two families of fibers, clearly evident from the discrete fiber simulations: the first family (red in Fig. 1) is aligned with the principal axes (shown by the white arrow in Fig. 1) and are in a state of tension while the second family of fibers (black, in compression) provide an isotropic background stress that opposes alignment. The stress at any material point is the sum of the stresses from these two components (Eq. 3). The degree of the interaction between the two families of fibers is determined by the parameter $E_f/E_b$ – when this ratio is large, the isotropic part provides little resistance to alignment. A systematic study of the range of force transmission as a function of this parameter is given below. With the two families of fibers, our model captures the key features of the response of a collagen network to force, in particular the knee and the subsequent hardening response.

## 3. Results

Having developed a constitutive law, we use it in analytical calculations and finite element simulations to study the impact of the material parameters of the isotropic and fibrous components of the matrix, the shape of cells and the polarization of cell contractile forces on force transmission in fibrous matrices. We have simulated cells on fibrous as well as linear and non-linear substrates to identify the key factors that allow for long range force transmission in fibrous matrices. All simulations were carried out using the finite element package Abaqus (24) by implementing the material model of the new fibrous constitutive law in a user material subroutine (details of the implementation are given in Appendix B). The numerical simulations were performed in a finite deformation setting (*i.e.* the effect of geometry changes on force balance and rigid body rotations are explicitly taken into account).

### 3.1 Force transmission in 3D matrices depends on the fibrous components and the magnitude of the contractile strains

To determine the impact of the fibrous component of the matrix on force transmission, we consider matrices that are linearly elastic, hyperelastic (neo-Hookean) and fibrous (characterized by the constitutive law Eq. 4). We consider the case of a spherical cell or contractile explant of radius $R$ in a 3D matrix contracting isotropically and inwardly by an amount $u_0$ (contractile strain $= u_0/R$) . In our calculations, we apply the boundary condition on the radial displacement $(= u_0)$ at the cell-matrix interface and determine the elastic fields in the matrix by applying both symmetry (or periodic) and fixed (all displacements and rotations vanish) conditions at the top and bottom surfaces of the matrix located at a distance $L \sim 10R$ from the center of the cell. In the case of the linearly elastic material, the scaled displacement fields $(u/u_0)$ are independent of the magnitude of the contractile strain, $u_0$, whereas this is not the case for non-linear materials. For both the neo-Hookean and the isotropic response of the fibrous material, the material parameters are chosen such that the Young's moduli and Poisson ratios are the same as that of the linear elastic material at small strains.

We find that the displacement fields decay rapidly within a distance on the order of the cell diameter in non-fibrous materials (Fig. 2a-black, blue and green curves), while the displacement fields are long ranged in the fibrous matrix (Fig. 2a- red and orange curves). The range of interaction in the fibrous matrix is more than 20× the radius of the cell as evidenced by the fact that the boundary condition (periodic vs. fixed) has an impact on the displacement fields; the cells in this case are able to feel its periodic image since the displacement field does not completely vanish at the boundaries.

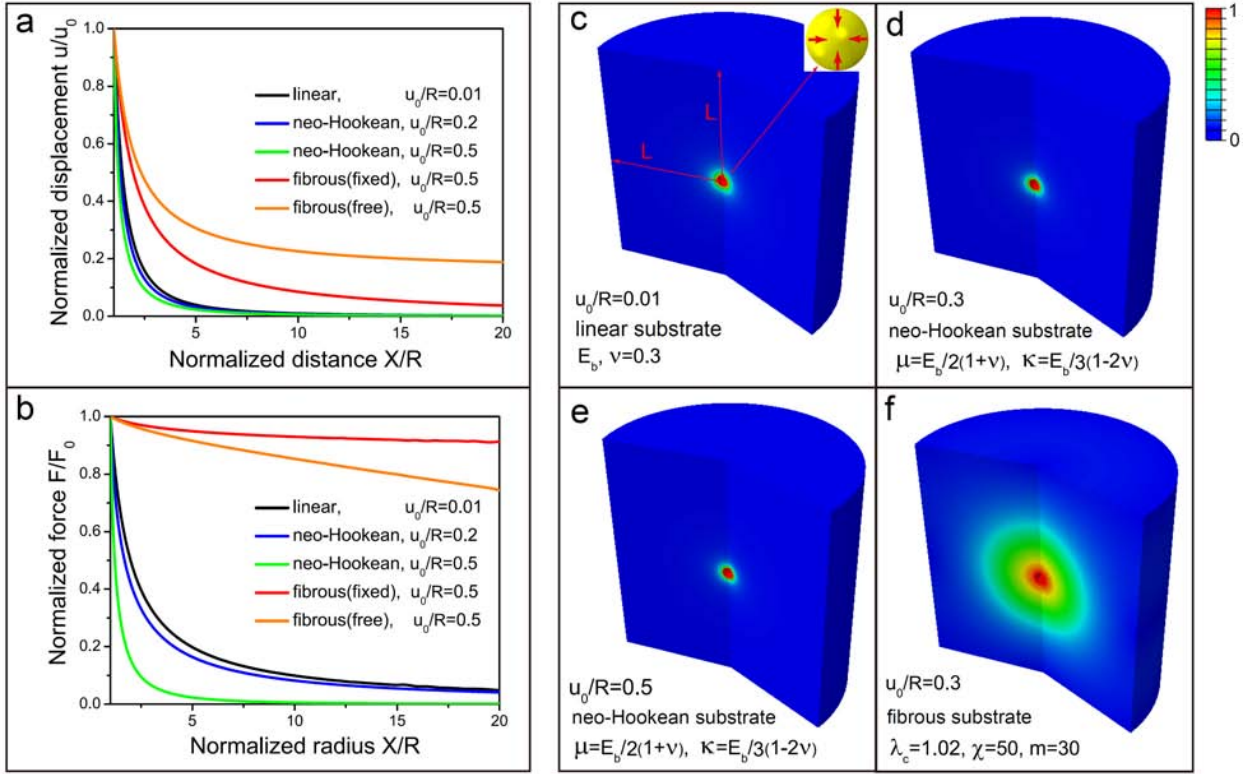

Figure 2: Displacement and force profiles in 3D linearly elastic, neo-Hookean and fibrous matrices with a spherical and isotropically contracting cell (radius = R). (a)-(b) Normalized radial displacement $u(X)/u(R)$ and force $F(X)/F(R)$ as functions of the normalized distance $X/R$ (the boundaries are located at a distance $L = 100R$) from the center. We have chosen the critical stretch, $\lambda_c = 1$, fibrous modulus $\chi = 50$ and the strain stiffening parameter $m = 0$ for the fibrous matrix; (c)-(f) Contour plot of normalized radial displacement $u(X)/u(R)$ for fibrous matrices with $\chi = 50$ and $m = 30$, $E_b = 2$kPa and Poisson's ratio $\nu = 0.3$ (same as $E_b$ and $\nu$ for linear matrices). For neohookean matrices $\mu/E_b = 1/2(1+\nu)$, $\kappa/E_b = 1/3(1-2\nu)$.

To gain further insight into the range of elastic fields, we plotted the total force $F(X) = \sigma(X)4\pi X^2$, normalized by the force at the cell-matrix interface, in Fig. 2b. We find that the decay of the total force in strain-hardening hyperelastic matrices is more rapid than in the case of the linearly elastic material, whereas the transmission of force is very long ranged in fibrous matrices. In Appendix C, we have derived a closed form expression for the decay of the force distribution as a function of the material parameters of the fibrous phase. These analytical calculations and the simulations in Fig. 2(c-f) clearly show that the fibrous components and not the isotropic strain-hardening response lead to long-range force transmission.

### 3.2 Force transmission in 3D matrices depends on the shape of cells or explants and cell polarization

Next, we consider the effect of shape and contraction anisotropy on force transmission in elastic and fibrous matrices. Unlike prior work that focused on the role of shape and cell polarization in linear elastic materials (25, 26), here we consider fibrous materials described by the constitutive laws derived in Sec. 2. We model elongated cells as prolate spheroids described by the shape,

$(x/a)^2+(y/a)^2+(z/b)^2=1$. Here $a$ and $b$ represent the length of the semi-minor and semi-major axes of the prolate spheroid, respectively. The polarization of active forces is modeled by assuming that the contractile strains (determined by molecular motors and regulation of adhesion sites) along the long axis of the spheroid, $\varepsilon_b$ are greater than the strain along the short axis, $\varepsilon_a$. In order to compare shapes with different aspect ratios, $\alpha = 1 - a/b$ and strain polarizations, $\beta = \left(1 - \frac{1-\varepsilon_b}{1-\varepsilon_a}\right)/(1 - \frac{V_1}{V_0})$, we assume that the volume of the cells prior to $(V_0 = 4\pi a^2 b/3 = 4\pi R^3/3)$ and after contraction $(V_1 = 4\pi(1-\varepsilon_a)^2(1-\varepsilon_b)R^3)$ are the same in all cases. Note that $\alpha = 0$ corresponds to a sphere, while $\alpha \sim 1$ is a highly elongated prolate spheroid. Similarly, $\beta = 0$ corresponds to isotropic contraction, while $\beta = 1$ represents a fully polarized cell (Fig. 3(a-d)). Here $R$ is the radius of the sphere as $\alpha = 0$. The above definitions also provide a definition for the size of a cell $R = a^{2/3}b^{1/3}$, which is the geometric mean of the lengths of the semi-major and semi-minor axes of the elongated cell (which can be considerably shorter than the length of the semi-major axis for a highly elongated cell).

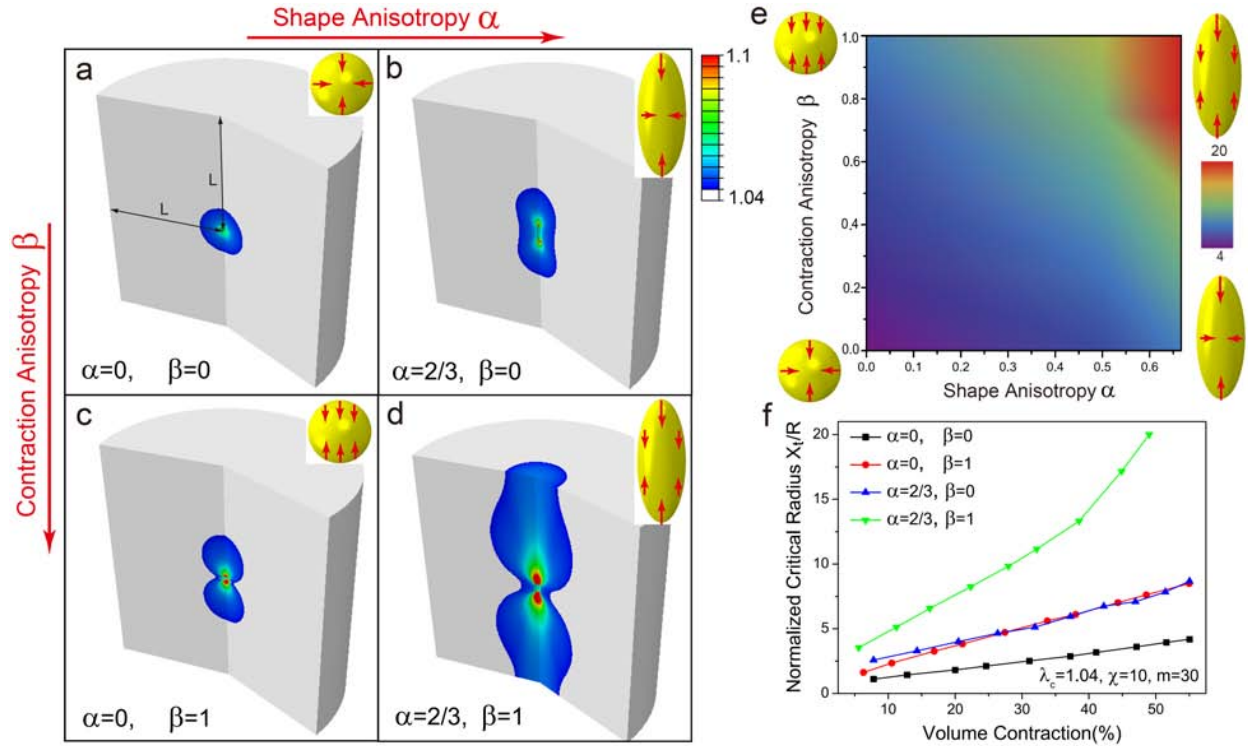

Figure 3: The influence of shape and contraction anisotropies ($\alpha$ and $\beta$, respectively) of contractile cells on distance $X_t/R$ over which forces are transmitted (measured by the extent of aligned fibrous regions in the matrices): (a)-(d) Contour plots of aligned (colored) and isotropic (white) regions for the 4 cases with $\alpha = 1, 2/3$, and $\beta = 0, 1$. Colors (from blue to red) represent maximum principle stretches (1.04-1.1); (e) Contour plots of $X_t/R$ as function of shape anisotropy $\alpha$ and contraction anisotropy $\beta$. Colors (from blue to red) represent $X_t/R$ $(4 - 20)$; (f) Normalized transmission distance $X_t/R$ vs volume contraction for the 4 cases in (a)-(d). Yellow ellipsoids with red arrows indicate contractile cells with different values of $\alpha$ and $\beta$. Material parameters for the fibrous matrix are $\lambda_c = 1.04, \chi = 50, m = 30$. The volume contraction is 55% for all the cases (a-d). The matrix size is $20 \times$ the contractile cell radius ($L/R = 20$) and the symmetry boundary conditions are applied at all boundaries.

The effect of shape and contraction and shape anisotropies on the range of force transmission in fibrous matrices is shown in Fig. 3 (a-d). Here the colored regions represent the extent of the aligned fibrous region, where the fibers are aligned with the tensile principal axis of strain tensor. We find that while shape and contraction anisotropy leads to an increase in the extent of the fibrous region, the effect is significantly amplified when both these factors are present simultaneously. We can understand this by noting that both shape and contraction anisotropy lead to concentration of tensile strains along the long axes of the cells. However, this effect is considerably magnified when the shape is elongated and the cell is polarized; significant concentration of tensile stresses in this case (Fig. 3d) leads to formation of extended regions where fibers are aligned. A heat map of the range of force transmission as a function of these parameters is given in Fig. 3e, for the case where the volume contraction is $1 - V_1/V_0 = 55\%$. We find that the extent of the fibrous region can be as high as 20× the characteristic size of the fully polarized cells for $\alpha = 2/3$, as has been observed by several groups (8, 18). The influence of the magnitude of volume contraction of the cell on the range of force transmission is plotted in Fig. 3f – our simulations show that range of force transmission generally increases with increase in overall contractile strain, although the effect is much more pronounced in elongated and polarized cells on account of the stress concentration effects discussed above. Thus, our analytical calculations and simulations collectively show that, in addition to the fibrous components of the matrix, elongated cells and polarized contraction leads to long-range force transmission.

**3.3 Long-range transmission in 3D matrices varies with the stiffness and strain-hardening exponent of the fibrous component and the critical strain for fiber formation**

We show in this section that in the material model that we have developed, the relative contributions of the fibrous and isotropic strain-hardening components to the overall mechanical response depends on three parameters: the ratio of the initial elastic moduli of the two components, $E_f/E_b$, the strain hardening exponent of the fibrous phase, m, and the critical strain for the onset of the fibrous response. A more pronounced fibrous response is obtained when $E_f/E_b$ and m are large and when the critical stretch, $\lambda_c$ is small (leading to an early transition to the aligned fiber phase). The extent of the aligned fibrous region that surrounds an elongated ($\alpha = 2/3$) and fully polarized ($\beta = 1$) cell is shown in Fig. 4(a-d) as a function of the material parameters that characterize the fibrous phase. The simulations show that the range of force transmission increases with increasing modulus and the strain hardening exponent of the fibrous phase and with decreasing values of the critical strain for transitioning to the fibrous phase. These parameters are determined by the density of fibers, the numbers of crosslinkers per fiber and the porosity of the fibrous gels as discussed in Sec. 2.

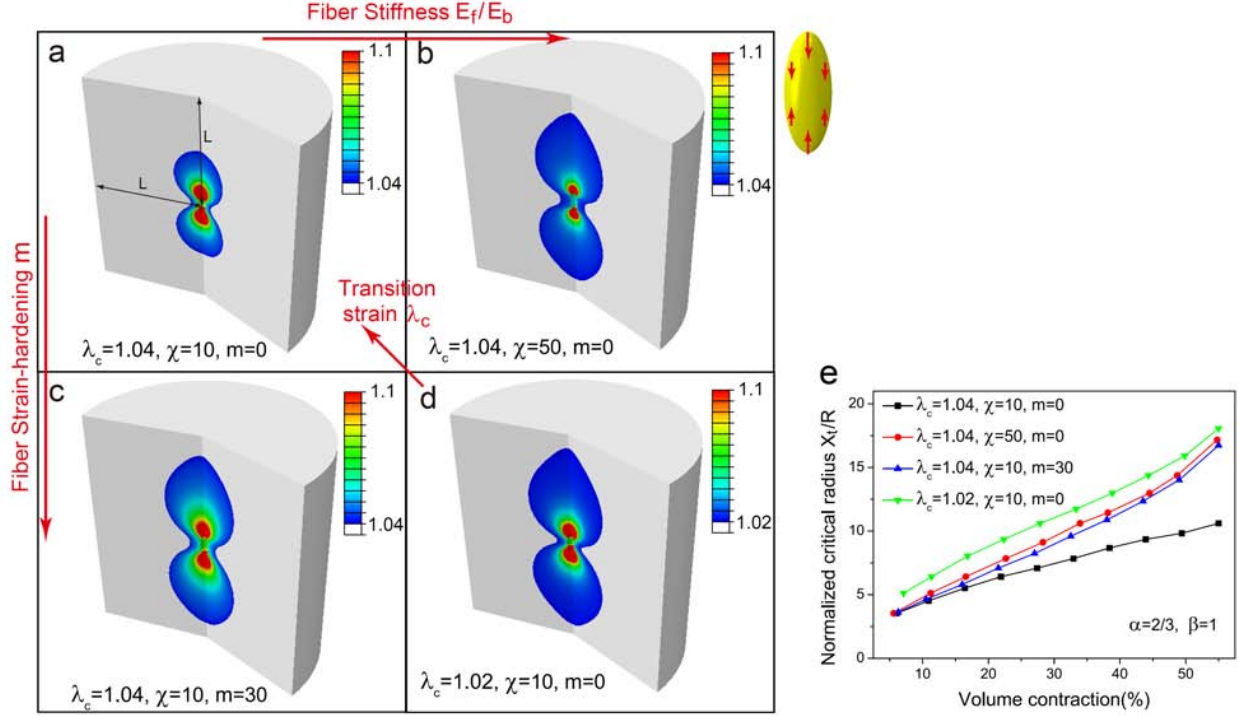

Figure 4: The influence of material parameters of fibrous matrices on the transmission distance $X_t/R$: (a)-(d) Contour plots of aligned (colored) and isotropic (gray) regions for the 4 cases with $\lambda_c = 1.02 - 1.04$, $\chi = 10 - 50$ and $m = 0 - 30$. Colors represent maximum principle stretch $(1.04 - 1.1)$ (increasing from blue to red). (e) Normalized transmission distance $X_t/R$ vs volume contraction for the 4 cases in (a)-(d). Shape and contraction anisotropies are $\alpha = 2/3$ and $\beta = 1$ and the volume contraction is 55% for all the cases. The matrix size is $20 \times$ the contractile cell radius ($L/R = 20$) and the symmetry boundary conditions are applied all boundaries.

## 3.4 Cells sense farther into fibrous substrates than into linear and strain hardening substrates

Recent work has demonstrated that fibroblasts sense deeper into collagen and fibrin gels (typically $> 65\ \mu m$) than they do into polyacrylamide gels (characteristic sensing distances of $< 5\ \mu m$) (18). In order to determine the characteristics of these gels responsible for characteristic sensing distances, we carried out calculations to determine cell sensing distance as a function of the thickness of gels constrained on one of the sides by a rigid (glass) substrate. Following earlier work (1), we assume the cell is circular and that it contracts radially inwards by pulling on the cell-substrate boundary. We apply displacement boundary conditions to this boundary (radial displacement $u(R)/R = 0.2$) and the bottom surface is clamped to the underlying glass substrate. All other surfaces are free of any traction. As in earlier work (1), we find that in both linear elastic and non-linear strain-hardening materials, the sensing distance is close to the radius of the cell, R. Increasing the gel thickness $H$ by a factor of 5 from $2/3\ R$ to $10/3R$ has very little impact on the spatial profiles of the displacement fields. On the other hand, cells are able to sense much deeper into fibrous gels as evidenced by the slower decay of the displacement fields of cells on thicker substrates. Our results for the sensing distances (Fig. 5g) show that cells on fibrous gels can sense up to 8× their radii compared to 1.8× the radii on strain-hardening substrates.

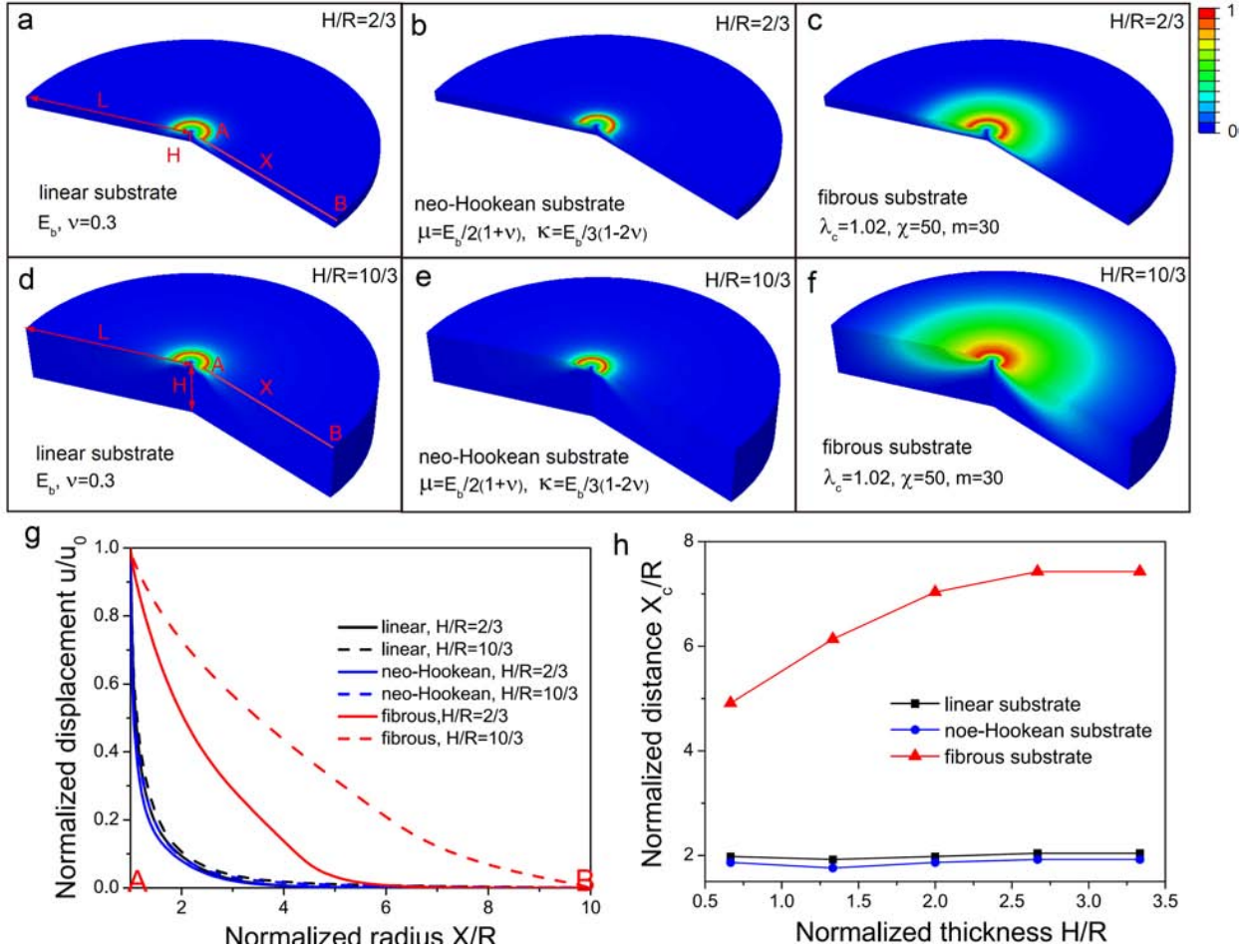

Figure 5: Mechanosensing distances for contractile cells on linear, neo-Hookean and fibrous substrates with thickness $H = 2/3 - 10/3R$, where R is the radius of the cell: (a)-(f) Contour plots of the normalized radial displacement $u(X)/u_0$ ($u_0 = u(R)$) with normalized thickness $H/R = 2/3$ (a)-(c) and $H/R = 10/3$ (d)-(f). (g) Normalized radial displacement $u(X)/u_0$ on the substrate surface as a function of the normalized distance $X/R$. (h) Normalized force transmission distance $X_c/R$ as a function of the normalized thickness $H/R$ (chosen with the criterion that the displacement fields decay by 90%, or $u(X_c)/u_0 = 0.1$). Circle (black), square (blue) and triangle (red) indicate linear, neo-Hookean and fibrous substrates, respectively. Material parameters for the fibrous matrix are $\lambda_c = 1.02, \chi = 50, m = 30$. The substrate radius is 10 × the cell radius ($L/R = 10$) and the bottom boundary is clamped.

### 3.5 Cells sense other cells located at distances ~20 times their size in fibrous 3D matrices

Interactions between pairs of cells play a key role in cell clustering during morphogenesis as well in pathological processes such as fibrosis, wound healing and metastasis. Based on our results (above) regarding the elastic fields of cells in different types of matrices, it is reasonable to guess that cell-cell interactions are significant when their separations are of the order of twice the sensing distance of a single cell. We verified this hypothesis by explicitly simulating the interactions between two cells in 3D fibrous and non-fibrous matrices as well as on substrates. The clear role of fibrous matrices in mediating cell-cell interactions is shown in Fig. 6 where significant overlap and alignment of strain-fields are observed for pairs of cells located in fibrous matrices at a distance of 10× their size. There is no overlap of strain fields for cells on non-fibrous substrates. Using these simulations, we confirm that cell-cell interactions become significant when cell spacing is twice the sensing distance, which is in agreement with the result in Fig. 6. Color represents the normalized radial displacement $(0 - 1)$ (increasing from blue to red). The geometry and boundary conditions of (b) and (e) are same as those in Fig 2f and Fig 5f, respectively. Our simulations also clearly show the formation of collagen lines observed experimentally between pairs of cells (4, 5); we find that that the alignment of fibers coincides with the line that connects the centers of the two contractile cells both in 3D matrices and on substrates (Fig. 6c and 6e). Thus, we find that fibrous but not neo-Hookean matrices enable cells to form collagen lines and interact mechanically with other cells at long range.

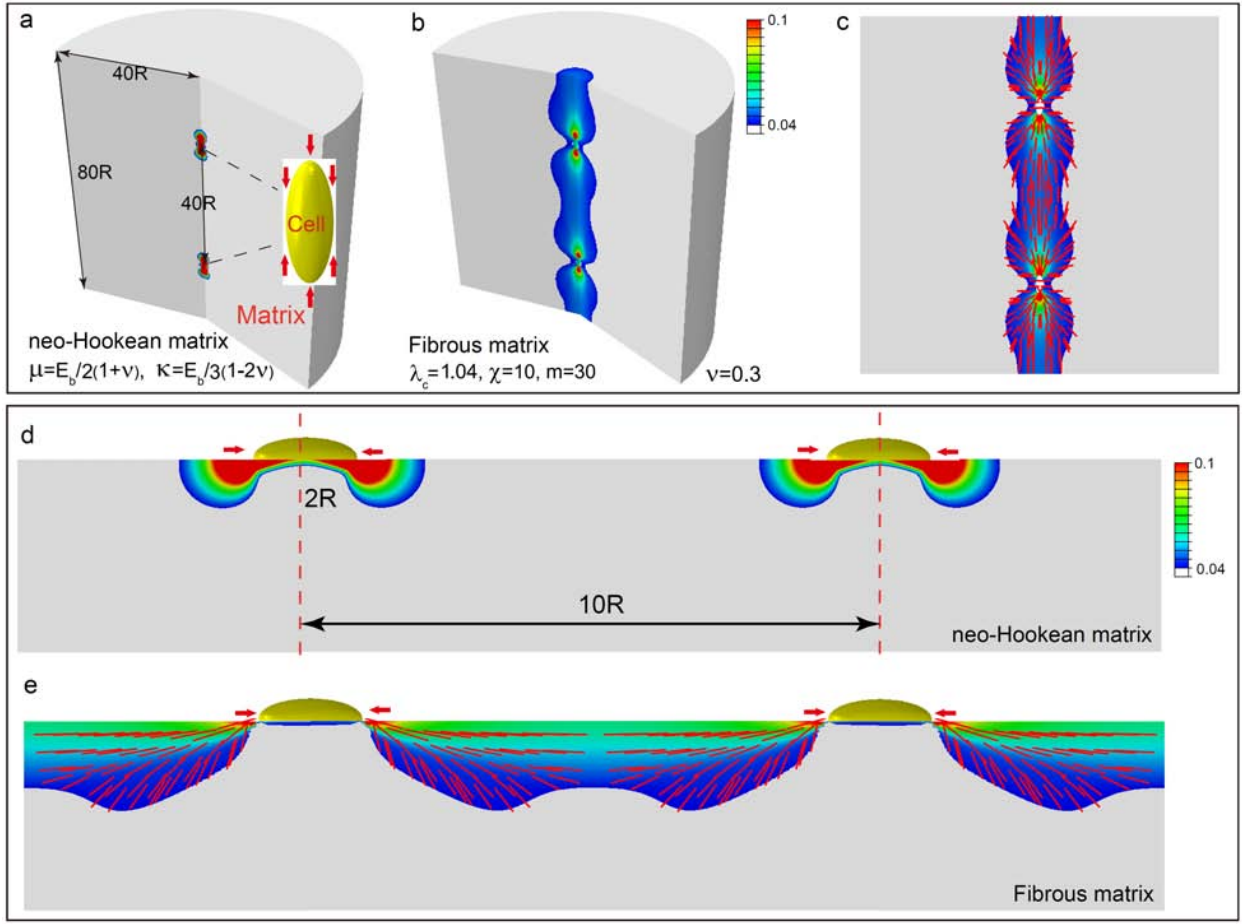

Figure 6: Interactions of pairs of contractile cells in neo-Hookean and fibrous matrices : (a-b) Contour plots of maximum principle strain in 3D matrices; (c) Vector plots of maximum principle strain (which coincides with the orientation of the collagen lines) in a 3D fibrous matrix; (d-e) Contour plots of maximum principle strain on 3D substrates. Colors (from blue to red) represent maximum principle strain (0.04-0.1). Lengths of red lines represent the magnitude of the maximum principle strain (0.04-1) and their orientations show the directions of fiber alignment. For the fibrous matrices, colored and gray regions represent aligned fibrous and isotropic regions, respectively. We have chosen $\lambda_c = 1.04$, $\chi = 50$, $m = 30$ for the fibrous material.

## 4. Summary and Discussion

In summary, we have developed a new constitutive law for fibrous matrices that predicts the following key cell behaviors:

1. Both shape and contraction anisotropy are important for long-range force transmission. These features of cells lead to stress concentration at the poles, which in turn leads to fiber alignment. Elongated prolate spheroidal cells with polarized contraction are able to sense the mechanical environment over much larger distances than spherical cells exhibiting diagonal contraction.

2. Tension-driven fiber alignment plays a crucial role in mechanosensing: small critical stretch for fiber alignment ($\lambda_c$), large fiber stiffness ($\chi$), and fiber strain hardening behavior ($m$) enable long-range interactions.

3. Cells in 3D fibrous matrices and cells on 2D fibrous substrates sense rigid boundaries and other cells over relatively long distances compared to cells in and on linear and strain-hardening isotropic materials. The range of force transmission increases with increasing contractility for cells in fibrous matrices while increasing contractility of cells cannot lead to enhancement of mechanosensing distances in linear and strain-hardening materials.

4. Cells in 3D fibrous matrices sense rigid boundaries over 10 × their diameters and other cells over 20 × their diameters. Cells on 2D fibrous substrates sense radial rigid boundaries up to 8 × their radii and thicknesses up to 3.5 × their radii. Sensing distances can be further enhanced by increasing cell elongation, polarization and contractility.

These findings are highly relevant biologically. They suggest that the presence of a fibrous matrix, as well as the material properties of that matrix, determine the nature of the mechanical interactions between groups of cells and between cells and boundaries in a range of settings including development, cancer metastases, and wound healing and fibrosis. This is consistent with the experimental observation that increased collagen cross-linking is associated with many of these processes, and suggests that studying the impact of other matrix proteins on fibrous collagen matrices may yield important insights into normal biology and pathology. Similarly, elongated cell shape and polarized cell contractility enhance long-range mechanical interactions; our results are consistent with experimental observations that cells involved in many of these processes are elongated and contractile (and may have undergone an epithelial to mesenchymal transition).

*Derivation of the constitutive law:* The constitutive law for fibrous matrices we have proposed is non-linear with respect to the orientation and the magnitudes of the principal strains. The direction of the stiffened fibrous response coincides with the principal orientations whose principal strains are above a critical threshold. As we show below, these two features are critical to capture the key features of long-range force transmission observed in experiments. In this regard, the detailed form of the constitutive law of the matrix is not crucial as along as it captures the orientational anisotropy and stiffening that naturally arises along the principal directions upon loading. We have verified this idea by studying force transmission in matrices (Fig. 7) with other functional forms of response (Appendix D), but those retain the general features of anisotropic stiffening that coincides with the principal strain orientations. In particular, our constitutive law shares some common features with modified Cauchy-Green deformation tensors (23, 27), but there are some crucial differences that are essential to obtaining long range force transmission. In this previously-published work, the collagen network is modeled as a hyperelastic material reinforced by two families of fibers whose orientations depend on the directions of principal stress (Appendix D). Note, however that unlike our formulation, their constitutive laws are based on the invariants of the modified Cauchy-Green deformation tensor. As we show in Appendix E, long-range force transmission cannot be observed when modified Cauchy-Green deformation tensors are used. We have therefore modified the constitutive law where we use the principal stretches, which are the eigenvalues of the Cauchy-Green deformation. We have, however, retained the feature that collagen fibers form only along those directions where the stretches are tensile. In essence, the law previously proposed relies on the deviatoric components of the dyadic $\mathbf{n}_i \otimes \mathbf{n}_i$ ($\mathbf{n}_i$ being the principal stretch), which as we show in Appendix E cannot give long-range force transmission since an incompressible material is similar to an isotropic material without tension-driven alignment of collagen fibers (Eq. C10 in Appendix C).

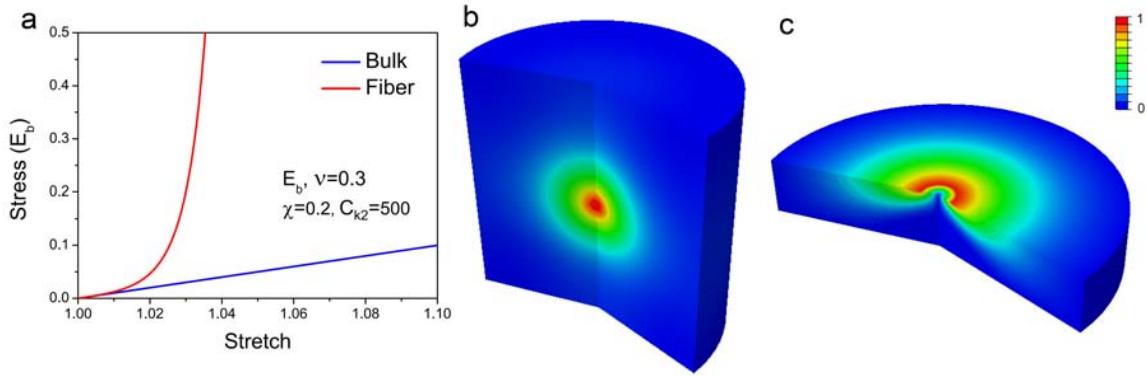

Figure 7: Force transmission for the material with strain energy function similar to that given in Ref. (23, 27) ($E_b, \nu = 0.3, \chi = 0.25, C_{k2} = 500$). (a) Blue and red curves represent bulk and fibrous contributions to the stress, respectively. (b) Contour plot of normalized radial displacement $u(X)/u(R)$ in fibrous matrices (which is similar to the result in Fig. 2f). (c). Contour plot of normalized radial displacement $u(X)/u(R)$ on a fibrous substrate, which is similar to the result in Fig. 5f. Colors (from blue to red) represent the normalized radial displacement ($0 - 1$). The geometry and boundary conditions for (b) and (c) are same as in Fig. 2f and Fig. 5f, respectively.

*Sensing of thickness and lateral boundaries by cells on substrates*: Our results are consistent with published experimental data on cell sensing distances. Both computational modeling (1, 2) and experimental observations (3, 28, 29) suggest that cells cultured on polyacrylamide (PA) gels (linear elasticity) cannot sense nearby cells beyond one cell length apart (<40 μm) (1) and substrate thickness beyond half a cell length away (<20 μm) (2). In contrast to cells on PA gels, human mesenchymal stem cells (hMSCs) and 3T3 fibroblasts on fibrin gels were shown to sense and respond to mechanical signals up to five cell lengths away (9), consistent with the results shown in Fig. 5f and 5g. Leong et al. studied the role of collagen I gel thickness on the fate of hMSCs and found that the mechanosensing distance for these cells is about 130 μm, which corresponds to approximately 4.3× cell radii, also in agreement with our work. Recently, Rudnicki et al. designed sloped collagen and fibrin gel cultures to investigate thickness sensing. They found human lung fibroblast (HLF) and 3T3 fibroblast cell areas gradually decrease as gel thickness increases from 0 to 150μm, with spreading affected on gels as thick as 150 μm (18). Since the spreading radius in the case of the 150 μm thick gel is 20 μm, the mechanosensing distance for substrate thickness is 7.5× cell radii (18). While these multiscale simulations suggest sensing distances of 3.7× cell radii (sensing distance of 50 µm for a cell radius of 13.4 µm), our results show that cells sense boundaries up to 3.5 × their radii on fibrous substrates compared to 1.8× their radii on strain-hardening substrates (Fig. 5h). Thus, our work provides a good estimate for sensing distances on fibrous substrates. While most of the experimental work has focused on thickness sensing, recently Mohammadi et al. developed a model system to examine sensing of lateral boundaries in floating thin collagen gels populated with 3T3 fibroblasts (30). They found that cell-induced deformation fields extended to, and were resisted by, the grid boundaries 250 µm away (30) suggesting a sensing distance for lateral rigid boundaries of about 8 × cell radii. These results are consistent with our calculations in Fig. 5g that show that both lateral and thickness sensing distances are similar in magnitude.

*Mechanosensing in 3D gels*: Our results are consistent with published experimental work on the importance of cell shape, cell contractility, contractile strains, and local fiber alignment on long-range force transmission. Gjorevsk and Nelson examined the strain fields around engineered 3D epithelial tissues in collagen I gels. They found that linear elasticity cannot explain the long-ranged nature of the strain fields but reported that mechanical heterogeneities caused by stiffening near the poles of elongated contractile epithelial tissues can explain the decay of strain fields (31). Our results clearly show that long-range displacement fields within matrices can be captured by tension-driven local fiber alignment, and that heterogeneities result from the anisotropic shape of the cell domain and the anisotropic contraction of cells (Fig 3). Cell contractility results in reorganization of the ECM to provide contact guidance that facilitates 3D migration and invasion (4, 5, 32). The fiber alignments observed between nearby cells in 3D matrices (4, 5, 32) are clearly shown in our FEM simulations (Fig 6c and 6e). Experimental work has shown that treatment of cells to abolish actomyosin contractility leads to dissolution of the collagen lines, in agreement with our results that show that the magnitude of contractile strains play an important role in determining the range of force transmission. Recent experiments on mammary acini in collagen gels show that they can interconnect by forming long collagen lines up to around 10 × acini size (7). Guo et al. find that these lines and interactions are initiated by traction forces created by the cells and not by diffusive factors (10). They also found collagen-density dependent transmission of force up to 10 × cell radii for interacting acini. Our results show that cells in 3D fibrous matrices can sense the radial rigid boundaries up to 10 × their diameters and the other cells up to 20 × their diameters (Figs. 3 and 6), which is in very good agreement with these experiments. Furthermore, Ma et al. suggest that the fibrous nature of the ECM leads to reorganization of the collagen fibers leading to areas of higher fiber density near the cells over relatively long distances (10 cell-diameters) (8). The mechanism whereby this reorganization proceeds (starting from a random network) is discussed in our work.

Koch et al. studied the effect of anisotropic cell shape and contractility on the range of force transmission in invasive and non-invasive cancer cells (19). They found that both lung and breast carcinoma cells were significantly elongated compared to the non-invasive cells, which were observed to have rounder shapes. Cell shape anisotropy was accompanied by a larger sensing distance, suggesting that directionality of traction forces is important for cancer cell invasion, consistent with our results (Fig. 3).

In sum, we present a new constitutive law that describes the behavior of cells in matrices. All of the parameters for our constitutive law can be obtained either from experiments or from fiber simulations as have been done in Fig. 1. Our findings are relevant to a variety of normal and pathological processes and, importantly, as highlighted in detail above, are consistent with an extensive body of experimental work. We hope that this work will inspire further experiments where the mechanical properties of the ECM are tuned by varying the fiber density and degree of crosslinking to validate our predictions.

## APPENDIX A: Discrete fiber simulations

We developed a finite element based 2D discrete fiber model that captures all aspects of network mechanics including non-affine stiffening, fiber alignment and bending-stretching transitions following our earlier work on crosslinked biopolymer networks (33). The 2D random fiber networks representing collagen gels are created with linear elastic fibers and rigid crosslinks (Fig. 1a). Fibers are uniformly distributed in the computational domain and a crosslink is formed when two fibers intersect. Collagen fibers have diameter in the range of few 100 nanometers to few microns and moduli of few 100 kPa (34–36). As the persistence length of collagen fibers is in the range of few microns, these fibers are typically modeled as linear elastic. Fibers are modeled using shear flexible Timoshenko beam elements in the finite element package, ABAQUS (24). Collagen gel considered in experiments is converted into a computational network (with equivalent fiber density) using the approach of Stein, Andrew M., et al (37). For the given concentration and volume of the gel, fiber radius is given by

$$r = \sqrt{\frac{V_g \rho_c v_c}{\pi L_{Tot}}}$$

where $V_g$ ($\mu m^3$) is the volume of the gel, $\rho_c (= 1 - 5\ mg/ml)$ is the mass density of collagen, $v_c = 0.73\ ml/g$ is the specific volume of collagen, $r$ ($\mu m$) is the radius of the fibers and $L_{Tot}$ ($\mu m$) is the total length of collagen in the gel. The 3D variables converted into equivalent the 2D ones by transforming quantities per unit volume to quantities per unit area. Fiber radius is assumed to be $250\ nm$ and from the above relation, the total length of fiber in the gel is calculated for varying collagen concentrations. The fibers have both flexural and stretching rigidities and the crosslinks are assumed to be rigid (38). A parametric study for various collagen concentrations ($2, 3, 4\ and\ 5\ mg/ml$), simulating simple shear deformation shows good agreement with the experimentally observed strain sweep results (39). Increasing gel concentration reduces the collagen mesh size (distance between two crosslinks) leading to a stiffer response. The reduction in the length of the fiber between the crosslinks affects the bending characteristics and leads to an increase in the initial stiffness and a decrease the knee strain.

## APPENDIX B: Finite element implementation of the fibrous constitutive law

All simulations were performed in a finite deformation setting. The matrices are modeled using 4-node bilinear axisymmetric quadrilateral elements. The axisymmetric constitutive law, the equilibrium condition, $\partial \sigma_{ij}/\partial dx_j = 0$, and the boundary conditions constitute a well-posed boundary value problem. We implemented the constitutive equation in a user material model in the finite element package ABAQUS (24). The tangent modulus tensor in the material description $\boldsymbol{C^{SC}}$, the tangent modulus tensor for the convected rate of the Kirchhoff stress $\boldsymbol{C^{\tau C}}$, the tangent modulus tensor for the Jaumann rate of the Kirchhoff stress $\boldsymbol{C^{\tau J}}$, and the material Jacobin $\boldsymbol{C^{MJ}}$ (needed for the user material model) can be expressed as (23, 40)

$$C^{SC}_{mnpq} = 4 \frac{\partial^2 W}{\partial C_{nm} \partial C_{pq}}$$

$$C_{ijkl}^{\tau C} = F_{im}F_{jn}F_{kp}F_{lq}C_{mnpq}^{SC} \tag{B1}$$

$$C_{ijkl}^{\tau J} = C_{ijkl}^{\tau C} + \delta_{ik}\tau_{jl} + \tau_{ik}\delta_{jl}$$

$$C_{ijkl}^{MJ} = C_{ijkl}^{\tau J}/J$$

Here the second Piola–Kirchhoff stress $\boldsymbol{\tau} = \boldsymbol{\sigma}/J$,

$$C_{ijkl}^{MJ} = C_{ijkl}^{b} + C_{ijkl}^{f}$$

$$C_{ijkl}^{b} = \frac{\mu}{J}\left(\frac{1}{2}(\delta_{ik}\bar{B}_{jl} + \bar{B}_{ik}\delta_{jl} + \delta_{il}\bar{B}_{jk} + \bar{B}_{il}\delta_{jk}) - \frac{2}{3}\delta_{ij}\bar{B}_{kl} - \frac{2}{3}\bar{B}_{ij}\delta_{kl} + \frac{2}{9}\delta_{ij}\delta_{kl}\bar{B}_{mm}\right) + \kappa(2J-1)\delta_{ij}\delta_{kl} \tag{B2}$$

$$\boldsymbol{C}^{f} = \frac{1}{J}\sum_{a=1}^{3}\frac{\partial}{\partial\lambda_{a}}\left(\frac{\partial f(\lambda_{a})}{\partial\lambda_{a}}\frac{1}{\lambda_{a}}\right)\lambda_{a}^{3}\boldsymbol{n}_{a}\otimes\boldsymbol{n}_{a}\otimes\boldsymbol{n}_{a}\otimes\boldsymbol{n}_{a}$$
$$+ \sum_{\substack{a,b=1\\a\neq b}}^{3}\frac{\sigma_{b}\lambda_{a}^{2} - \sigma_{a}\lambda_{b}^{2}}{\lambda_{b}^{2} - \lambda_{a}^{2}}(\boldsymbol{n}_{a}\otimes\boldsymbol{n}_{b}\otimes\boldsymbol{n}_{a}\otimes\boldsymbol{n}_{b} + \boldsymbol{n}_{a}\otimes\boldsymbol{n}_{b}\otimes\boldsymbol{n}_{b}\otimes\boldsymbol{n}_{a}) + \boldsymbol{I}\bar{\otimes}\boldsymbol{\sigma}^{f} + \boldsymbol{\sigma}^{f}\bar{\otimes}\boldsymbol{I}$$

Here we have adopted the abbreviations $(A\otimes B)_{ijkl} = A_{ij}B_{kl}$ and $(A\bar{\otimes}B)_{ijkl} = A_{ik}B_{jl}$. We define

$$\sigma_{a} = \frac{1}{J}\frac{\partial f(\lambda_{a})}{\partial\lambda_{a}}\lambda_{a} \tag{B3}$$

If $\lambda_b \to \lambda_a$, $\frac{\sigma_b\lambda_a^2 - \sigma_a\lambda_b^2}{\lambda_b^2 - \lambda_a^2}$ gives us 0/0 and must be determined using the limiting conditions (23),

$$\lim_{\lambda_b \to \lambda_a}\frac{\sigma_{b}\lambda_{a}^{2} - \sigma_{a}\lambda_{b}^{2}}{\lambda_{b}^{2} - \lambda_{a}^{2}} = \frac{1}{2}\frac{d\sigma_{a}}{d\lambda_{a}}\lambda_{a} - \sigma_{a} \tag{B4}$$

Integrating Eq. 4, the energy function $f(\lambda_a)$ can be expressed as,

$$f(\lambda_a) = \begin{cases} 0, & \lambda_a < \lambda_1 \\ \dfrac{E_f(\frac{\lambda_a - \lambda_1}{\lambda t})^n(\lambda_a - \lambda_1)^2}{8(1+n)(2+n)}, & \lambda_1 \leq \lambda_a < \lambda_2 \\ E_f\left[\begin{array}{l}-\dfrac{1}{2+3m+m^2} - \dfrac{\lambda_a}{1+m} + \dfrac{(1+\lambda_a-\lambda_2)^{2+m}}{(1+m)(2+m)} + \dfrac{\lambda_2}{1+m} \\ + \dfrac{(\lambda_a - \lambda_2)(\lambda_2 - \lambda_1)}{1+n} + \dfrac{4(\lambda_2 - \lambda_c)^2}{2+3n+n^2}\end{array}\right], & \lambda_a \geq \lambda_2 \end{cases} \tag{B5}$$

The second derivative of Eq. 4 can be expressed as,

$$\frac{\partial^2 f(\lambda_a)}{\partial \lambda_a^2} = \begin{cases} 0, & \lambda_a < \lambda_1 \\ E_f \left(\frac{\lambda_a - \lambda_1}{\lambda_2 - \lambda_1}\right)^n, & \lambda_1 \leq \lambda_a < \lambda_2 \\ E_f (1 + \lambda_a - \lambda_2)^m, & \lambda_a \geq \lambda_2 \end{cases} \quad (B6)$$

Here $\lambda_1 = \lambda_c - \lambda_t/2$, $\lambda_2 = \lambda_c + \lambda_t/2$.

**APPENDIX C: Analytical linear solution for the spherically symmetric case**

We further introduce Green-Lagrange strain tensor $\boldsymbol{\varepsilon} = (\boldsymbol{C} - \boldsymbol{I})/2$. For infinitesimal strains $\boldsymbol{\varepsilon}$ with $|\varepsilon_{ij}| \ll 1$,

$$J = 1 + \text{tr}(\boldsymbol{\varepsilon})$$
$$\bar{\boldsymbol{B}} = \boldsymbol{I} + 2\boldsymbol{\varepsilon} \quad (C1)$$
$$\lambda_a = (1 + 2\varepsilon_a)^{1/2} = 1 + \varepsilon_a$$

Substituting Eq. C1 into Eq. 2

$$\boldsymbol{\varepsilon} = \sum_{a=1}^{3} \varepsilon_a \boldsymbol{n_a} \otimes \boldsymbol{n_a} \quad (C2)$$

The fiber energy function in Eq. 1 can also be expressed as $f(\lambda_a) = U(\varepsilon_a)$,

$$\frac{\partial f(\lambda_a)}{\partial \lambda_a} = \frac{\partial U(\varepsilon_a)}{\partial \varepsilon_a} \frac{\partial \varepsilon_a}{\lambda_a} = \frac{\partial U(\varepsilon_a)}{\partial \varepsilon_a} \quad (C3)$$

Substituting Eq. C3 into Eq.3, we get

$$\boldsymbol{\sigma} = \boldsymbol{\sigma}^b + \boldsymbol{\sigma}^f,$$
$$\boldsymbol{\sigma}^b = \kappa \, \text{tr}(\boldsymbol{\varepsilon})\boldsymbol{I} + 2\mu \, \text{dev}(\boldsymbol{\varepsilon}), \quad (C4)$$
$$\boldsymbol{\sigma}^f = \sum_{a=1}^{3} \frac{\partial U(\varepsilon_a)}{\partial \varepsilon_a} \boldsymbol{n_a} \otimes \boldsymbol{n_a}$$

For linear bulk and fibrous response ($\lambda_c = 1$ and $m = 0$ in Eq. 4), Eq. C4 can be rewritten as,

$$\boldsymbol{\sigma} = \boldsymbol{\sigma}^b + \boldsymbol{\sigma}^f$$
$$\boldsymbol{\sigma}^b = \frac{E_b}{3(1 - 2\nu)} \text{tr}(\boldsymbol{\varepsilon})\boldsymbol{I} + \frac{E_b}{1 + \nu} \text{dev}(\boldsymbol{\varepsilon}) \quad (C5)$$
$$\boldsymbol{\sigma}^f = \sum_{a=1}^{3} E_f \, \boldsymbol{n_a} \otimes \boldsymbol{n_a}.$$

For infinitesimal strains, we have the geometric relations,

$$\varepsilon_r = \frac{du}{dr}, \qquad \varepsilon_\theta = \varepsilon_\varphi = \frac{u}{r}, \qquad J = 1, \tag{C6}$$

Here $u$ is the radial displacement and the constitutive law Eq. C5 can be rewritten as,

$$\sigma_r = \frac{E_b}{(1-2v)(1+v)}\left[(1-v)\frac{du}{dr} + 2v\frac{u}{r}\right] + E_f \frac{du}{dr} \tag{C7}$$

$$\sigma_\theta = \sigma_\varphi = \frac{E_b}{(1-2v)(1+v)}\left(\frac{u}{r} + v\frac{du}{dr}\right)$$

The condition for mechanical equilibrium $\frac{d\sigma_r}{dr} + \frac{2}{r}(\sigma_r - \sigma_\theta) = 0$ can then be written as,

$$\left[1 + \frac{(1+v)(1-2v)}{(1-v)}\frac{E_f}{E_b}\right]\left(\frac{d^2u}{dr^2} + \frac{2}{r}\frac{du}{dr}\right) - 2\frac{u}{r^2} = 0 \tag{C8}$$

The boundary condition is

$$u(r_0) = u_0, \quad u(\infty) = 0 \tag{C9}$$

The solution is

$$u(r)/u_0 = (r_0/r)^n \tag{C10}$$

$$\sigma_r(r)/\sigma_r(r_0) = (r_0/r)^{n+1}$$

Here $n = \frac{1}{2}\left(\sqrt{\frac{9+\chi}{1+\chi}} + 1\right)$ and $\chi = \frac{(1+v)(1-2v)}{(1-v)}\frac{E_f}{E_b}$

The strains and stresses can then be expressed as

$$\varepsilon_r = n\frac{u_0}{r_0}\left(\frac{r_0}{r}\right)^{n+1} \tag{C11}$$

$$\varepsilon_\theta = \varepsilon_\varphi(r) = -\frac{u_0}{r_0}\left(\frac{r_0}{r}\right)^{n+1}$$

$$\sigma_r = \left\{\frac{E_b}{(1+v)(1-2v)}[(1-v)n - 2v] + nE_f\right\}\frac{u_0}{r_0}\left(\frac{r_0}{r}\right)^{n+1}$$

$$\sigma_\theta = \sigma_\varphi = \frac{E_b}{(1+v)(1-2v)}[vn - 1]\frac{u_0}{r_0}\left(\frac{r_0}{r}\right)^{n+1}$$

In the limit of strong fibrous response, $E_f/E_b \gg 1$, we find that the exponent $n \to 1$, whereas for an isotropic material for which $E_f/E_b \ll 1$, we find that $n \to 2$. Thus, stresses decay less precipitously, leading to an increased zone of influence in fibrous materials. This result is also consistent with theoretical estimates by Sander (41), who considered a less general case, $E_f/E_b \gg 1$, without including the effect of the Poisson's ratio, $v$.

**APPENDIX D: Strain energy function with the modified right Cauchy–Green tensor**

Holzapfel et al. (23, 27) developed a constitutive law to describe the mechanical response of arterial tissue with a strain energy function

$$W_f = W_b(\bar{I}_1, J) + \overline{W_f}(\overline{C}) = W_b(\bar{I}_1, J) + \sum_{i=4,6} f_i(\bar{I}_i) \tag{D1}$$

$$f = \begin{cases} 0, \bar{I}_i < 1 \\ \frac{k_1}{2k_2}\{\exp[k_2(\bar{I}_i - 1)^2] - 1\}, \bar{I}_i \geq 1, \end{cases}$$

where the first term $W_b$ represents the isotropic bulk response of the matrix (same as our model) and the second term $\overline{W_f}$ represents anisotropic stiffening due to two families of reinforcing collagen fibers that evolve during loading. The modified right Cauchy–Green tensor is $\overline{C} = C/J^{2/3}$. $\bar{I}_1$, $\bar{I}_4$ and $\bar{I}_6$ are the modified invariants of $\overline{C}$, which represent the squares of the stretches along the two families of fibers,

$$\bar{I}_1 = \mathrm{tr}(\overline{C}) \qquad \bar{I}_4 = N_4 \overline{C} N_4 \qquad \bar{I}_6 = N_6 \overline{C} N_6 \tag{D2}$$

where $N_4$ and $N_6$ are the unit vectors along the fibers in the reference configuration. Then, the Cauchy stress has the form,

$$\boldsymbol{\sigma} = \boldsymbol{\sigma}^b + \boldsymbol{\sigma}^f = \boldsymbol{\sigma}^b + \sum_{i=4,6} 2 \frac{\partial f_i(\bar{I}_i)}{\partial \bar{I}_i} \mathrm{dev}(\boldsymbol{n}_i \otimes \boldsymbol{n}_i) \tag{D3}$$

where $\boldsymbol{n}_4 = \boldsymbol{F} N_4$ and $\boldsymbol{n}_6 = \boldsymbol{F} N_6$ are the fiber vectors in the current configuration:

$$\boldsymbol{n}_4 = \boldsymbol{F} N_4, \qquad \boldsymbol{n}_6 = \boldsymbol{F} N_6 \tag{D4}$$

An iterative procedure starting with an arbitrary configuration of the fibers is implemented to find the fiber vectors in the reference and current configurations, $N_4$ and $\boldsymbol{n}_4$. By considering this constitutive law for the case of spherically-symmetric contractile strain, we show in Appendix E that this constitutive law cannot show long-range transmission of forces.

To enable the long range formation in fibrous media, the above strain energy function for collagen fiber alignment can be modified by using a Cauchy-Green deformation tensor instead of a modified Cauchy-Green deformation tensor. Denoting the principal stretches by $\lambda_a$, we retain the functional form of the function, $f(\lambda_a)$, such that it vanishes when the principal stretches are negative to get

$$f(\lambda_a) = \begin{cases} 0, \lambda_a < 1 \\ \frac{C_{k1}}{2C_{k2}}[\exp(C_{k2}(\lambda_a^2 - 1)^2) - 1], \lambda_a \geq 1 \end{cases} \tag{D5}$$

$$\frac{\partial f(\lambda_a)}{\partial \lambda_a} = \begin{cases} 0, \lambda_a < 1 \\ 2C_{k1}\exp(C_{k2}(\lambda_a{}^2 - 1)^2)(\lambda_a{}^2 - 1)\lambda_a, \lambda_a \geq 1 \end{cases} \quad (D6)$$

$$\frac{\partial^2 f(\lambda_i)}{\partial \lambda_a{}^2} = \begin{cases} 0, \lambda_a < 1 \\ 2C_{k1}\exp(C_{k2}(\lambda_a{}^2 - 1)^2)[4C_{k2}\lambda_a{}^6 - 8C_{k2}\lambda_a{}^4 + (3 + 4C_{k2})\lambda_a{}^2 - 1], \lambda_a \geq 1 \end{cases} \quad (D7)$$

Here $C_{k1}$ and $C_{k1}$ are the parameters for initial stiffness and strain-hardening. Note that $\bar{I}_i$ in the original form is replaced with $I_i$. We set $\chi = (1 + \nu)(1 - 2\nu)C_{k1}/(1 - \nu)E_b = 0.2$ and $C_{k2} = 500$ in our numerical simulations (Fig. 7).

**APPENDIX E: Analytical solution for the constitutive law with the modified right Cauchy–Green tensor**

Consider the special case of a spherical cell with isotropic contraction embedded in a fibrous matrix. As in the case of linear analysis in Appendix B, the deviatoric constitutive law in Eq. D3 can be rewritten for infinitesimal strains,

$$\boldsymbol{\sigma}^b = \kappa \, \text{tr}(\boldsymbol{\varepsilon}) \, \boldsymbol{I} + 2\mu \boldsymbol{e} + \sum_{i=4,6} \frac{\partial U(e_i)}{\partial e_i} \text{dev}(\boldsymbol{n}_i \otimes \boldsymbol{n}_i) \quad (E1)$$

Here the fiber energy function can be express as $f(\bar{I}_i) = U(e_i)$ with $\bar{I}_i = 1 + 2e_i$. For spherical symmetry, the deviatoric strain $e_r = \frac{2}{3}(\varepsilon_r - \varepsilon_\theta) \geq 0$ and $e_\theta = e_\varphi = \frac{1}{3}(\varepsilon_\theta - \varepsilon_r) \leq 0$, so Eq. E1 can be rewritten as,

$$\sigma_r = \frac{E_b}{3(1 - 2\nu)}(\varepsilon_r + 2\varepsilon_\theta) + \frac{2}{3}[\frac{E_b}{(1 + \nu)} + E_f](\varepsilon_r - \varepsilon_\theta) \quad (E2)$$

$$\sigma_\theta = \frac{E_b}{3(1 - 2\nu)}(\varepsilon_r + 2\varepsilon_\theta) - \frac{1}{3}[\frac{E_b}{(1 + \nu)} + E_f](\varepsilon_r - \varepsilon_\theta)$$

Using the relations $\varepsilon_r = \frac{du}{dr}, \varepsilon_\theta = \varepsilon_\varphi = \frac{u}{r}$ \quad (E3)

and the condition for mechanical equilibrium,

$$\frac{d\sigma_r}{dr} + \frac{2}{r}(\sigma_r - \sigma_\theta) = 0 \quad (E4)$$

we get

$$\frac{d^2 u}{dr^2} + \frac{2}{r}\frac{du}{dr} - 2\frac{u}{r^2} = 0 \quad (E5)$$

From boundary conditions: $u(r_0) = u_0, u(\infty) = 0$, the solution of Eq. E5 is

$$u(r)/u_0 = (r_0/r)^2 \quad (E6)$$

$$\sigma_r(r)/\sigma_r(r_0) = (r_0/r)^3$$

Comparing this with Eq. C 10, we find that the constitutive law of Holzapfel et al. (23, 27) does not show long range force transmission.

**Acknowledgements**

We thank Tom Lubensky for insightful discussions on fibrous constitutive laws. Research reported in this publication was supported by the National Institute of Biomedical Imaging and Bioengineering of the National Institutes of Health under Award Number R01EB017753 and the US National Science Foundation Grant CMMI-1312392. The content is solely the responsibility of the authors and does not necessarily represent the official views of the National Institutes of Health or the National Science Foundation.